\documentstyle[12pt]{article}
\begin{document}

\newcommand{\be}{\begin{equation}}
\newcommand{\ee}{\end{equation}}
\newcommand{\ben}{\begin{eqnarray}
\displaystyle}
\newcommand{\een}{\end{eqnarray}}

\newcommand{\la}{{\lambda}}
\newcommand{\Si}{{\Sigma}}
\newcommand{\de}{{\delta}}
\newcommand{\tde}{{\tilde \delta}}
\newcommand{\bde}{{\bar \delta}}

\newcommand{\C}{{\cal C}}
\newcommand{\cP}{{\cal P}}
\newcommand{\cA}{{\cal A}}
\newcommand{\cK}{{\cal K}}
\newcommand{\cD}{{\cal D}}
\newcommand{\cM}{{\cal M}}
\newcommand{\cJ}{{\cal J}}
\newcommand{\cJn}{{\cal J}_{\infty}}
\newcommand{\cO}{{\cal O}}
\newcommand{\cQ}{{\cal Q}}
\newcommand{\cR}{{\cal R}}
\newcommand{\cS}{{\cal S}}
\newcommand{\cT}{{\cal T}}
\newcommand{\cH}{{\cal H}}
\newcommand{\cY}{{\cal Y}}
\newcommand{\cZ}{{\cal Z}}
\newcommand{\cL}{{\cal L}}

\newcommand{\p}{\partial}
\newcommand{\na}{\nabla}
\newcommand{\hna}{\hat \nabla}
\newcommand{\hD}{\hat D}
\newcommand{\ints}{\int_{\Sigma} d\Sigma}
\newcommand{\LieN}{{\cal L}_{N^{i}}}
\newcommand{\Lief}{{\cal L}_{\phi^{i}}}
\newcommand{\Liet}{{\cal L}_{t^{i}}}
\newcommand{\LieM}{{\cal L}_{M^{\mu}}}
\newcommand{\Lie}{{\cal L}}

\newcommand{\tiA}{{\tilde A}}
\newcommand{\tiB}{{\tilde B}}
\newcommand{\tP}{{\tilde P}}
\newcommand{\tTheta}{{\tilde \Theta}}
\newcommand{\tim}{{\tilde \mu}}
\newcommand{\tir}{{\tilde r}}
\newcommand{\trp}{{\tilde r_{+}}}
\newcommand{\hr}{{\hat r}}
\newcommand{\rv}{{r_{v}}}
\newcommand{\dr}{{d \over d \hr}}
\newcommand{\epk}{{(\ep^{I})^\dagger}}

\newcommand{\bphi}{{\bar \phi}}
\newcommand{\balpha}{{\bar \alpha}}
\newcommand{\bbeta}{{\bar \beta}}
\newcommand{\brho}{{\bar \rho}}
\newcommand{\bbe}{{\bar \eta}}
\newcommand{\bmu}{{\bar \mu}}
\newcommand{\ab}{\underline A}
\newcommand{\apb}{\underline A'}
\newcommand{\bb}{\underline B}
\newcommand{\bbb}{\underline B'}
\newcommand{\bJ}{{\bar J}}
\newcommand{\bz}{{\bar z}}
\newcommand{\hhZ}{{\hat Z}}
\newcommand{\hhS}{{\hat \Sigma}}
\newcommand{\hhD}{{\hat D}}
\newcommand{\tF}{{\tilde F}}
\newcommand{\hB}{\hat B}
\newcommand{\bo}{\bar o}
\newcommand{\bm}{\bar m}
\newcommand{\bi}{\bar i}
\newcommand{\bX}{\bar X}
\newcommand{\bY}{\bar Y}
\newcommand{\bsi}{\bar \sigma}
\newcommand{\bla}{\bar \lambda}
\newcommand{\bdelta}{\bar \delta}
\newcommand{\bga}{\bar \gamma}
\newcommand{\ho}{\hat 0}
\newcommand{\hp}{\hat 5}
\newcommand{\hje}{\hat 1}
\newcommand{\hd}{\hat 2}
\newcommand{\httt}{\hat 3}
\newcommand{\tT}{\tilde T}
\newcommand{\hC}{\hat C}

\newcommand{\ep}{\epsilon}
\newcommand{\bep}{\bar \epsilon}
\newcommand{\Ga}{\Gamma}
\newcommand{\om}{\omega}
\newcommand{\si}{\sigma}
\newcommand{\ga}{\gamma}
\newcommand{\hth}{\hat \theta}
%%%%%%%%%%%%%%%%%%%%%%%%%%%%%%%%%%%%%%%%%%%%%%%%%%%%%%%%%%%%%%%%%%%%%%

\def\eth{{\mathord{\mkern8mu\bf\mathaccent"7020{\mkern-8mu\partial}}}}
\def\barm{{\overline m}}

%%%%%%%%%%%%%%%%%%%%%%%%%%%%%%%%%%%%%

\title{Positivity of Energy in Einstein-Maxwell
Axion-dilaton Gravity}
\author{Marek Rogatko\\
%\thanks{
%Supported in part by KBN grant.} \\
Institute of Physics \\
Maria Curie-Sklodowska University \\
20-031 Lublin, pl.Marii Curie Sklodowskiej 1, Poland \\
rogat@tytan.umcs.lublin.pl \\
rogat@kft.umcs.lublin.pl}

\date{\today}

\maketitle
%%%%%%%%%%%%%%%%%%%%%%%%%%%%%%%%%%%%%%%%%%%%%%%%%%%%%%%%%%%%%%%%%%

%%%%%%%%%%%%%%%%%%%%%%%%%%%%%%%%%%%%%%%%%%%%%%%%%%%%%%%%%%%%%%%%%%
\begin{abstract}
Avoiding the problem of the existence of asymptotically constant
spinors satisfying certain differential equations on a non-compact
hypersurface we presented the proof of positivity of the ADM
and Bondi energy in Einstein-Maxwell axion-dilaton gravity. In our 
attitude spinor fields defining the energy need only be defined near
infinity and there satisfying propagation equations.
\end{abstract}
%%%%%%%%%%%%%%%%%%%%%%%%%%%%%%%%%%%%%%%%%%%%%%%%%%%%%%%%%%%%%%%%%%%%

\section{Introduction}
In general relativity there are two distinct notions of
energy-momentum for an isolated system. The Arnowitt-Deser-Misner
(ADM) four momentum \cite{adm} defined at spatial infinity and the
Bondi one \cite{bbb}
measured at null infinity. During last years there were heavy attempts
to prove positivity of energy in general relativity.
The first complete proof was presented by Schoen and Yau \cite{sch}
and then afterwards by Witten \cite{wit}. Parker and Taubes \cite{par}
gave mathematically rigorous proof of the positive energy 
theorem analyzing the
falloff behaviour of spinors near infinity. Then Nester \cite{nes}
treated the problem of the positive mass theorem in a fully covariant
way to avoid difficulties concerning three-dimensional truncation of
the four-dimensional divergence theorem. In several works \cite{bon}
the issue of extending the Witten's method to the Bondi energy was
achieved. 
\par
Similar techniques were used to prove extensions of these results 
to the case of Einstein-Maxwell (EM) theory \cite{em}, supergravity
\cite{sup} and Kaluza-Klein theory \cite{kk}. It was shown 
\cite{gib} 
that self-gravitating solitons saturated a gravitational version
of the Bogomolnyi bound on energy due to
inclusion of Yang-Mills and Yang-Mills-Higgs and dilaton interactions.
In the case of the low-energy string theory,
the so-called Einstein-Maxwell axion-dilaton gravity (EMAD),
the results of positivity
of the ADM and Bondi energy were achieved in \cite{rog1,rog2}. 
The other problem
was to extend the positive mass theorem to asymptotically flat
manifolds containing black holes \cite{bmas}.\\
But one of the major problem connected with the Witten's type 
of proofs is
the existence of solutions to an elliptic system of partial
differential equations on a non-compact hypersurface. 
If the hypersurface is compact we can refer to the known theorems
for existence but in the case of non-compactness it should be necessary
to device new mathematical theorems (perhaps quite complicated).
In \cite{ber}
it was shown that positivity of the ADM and Bondi energy in
Einstein gravity could be achieved in a slight different way, without
proving the existence theorems on non-compact hypersurfaces.
\par
In our paper we generalize and
extend the reasoning presented in \cite{ber} to the case
of EMAD gravity.
We shall use two-component spinor notation \cite{pen}. 
Spinor indices will be denoted by capital letters. Our signature of 
$g_{\mu \nu}$ is $(+ - - -)$ and convention for the curvature tensor is
$2 \na_{[ \alpha } \na_{\beta ]} \eta_{\ga} = - R_{\alpha \beta \ga
\delta}\eta^{\delta}$. The Einstein equations are $G_{\mu \nu}
= - T_{\mu \nu}$.

%%%%%%%%%%%%%%%%%%%%%%%%%%%%%%%%%%%%%%%%%%%%%%%%%%%%%%%%%%%%%%%%%%%%%%
\section{Positivity of energy in EMAD gravity}
The low-energy limit of 
the heterotic string theory compactified on a six-dimensional
torus $T^6$ consists of the pure $N = 4,~ d = 4$ supergravity coupled to $N
= 4$ super Yang-Mills. One can truncate consistently this theory to
the simpler pure supergravity one. The truncation is based on
introducing always equal numbers of Kaluza-Klein and winding number
modes for each cycle. The truncated theory still exhibits $S$ and $T$
dualities. Therefore $N = 4,~ d = 4$ supergravity provides a simple
framework for studying classical solutions which can be considered
as solutions of the full effective string theory. The bosonic sector of
this theory with a simple vector field is called in literature EMAD
gravity. 
EMAD gravity provides after all a non-trivial generalization of
EM gravity, which consists of
coupled system 
containing a metric $g_{\mu \nu}$, $U(1)$ vector fields
$A_{\mu}$, a dilaton $\phi$ and three-index antisymmetric tensor field
$H_{\alpha \beta \ga}$. The action has the form \cite{ka}
\be
I = \int d^4 x \sqrt{-g} 
\left [ - R + 2(\na \phi)^{2} + {1 \over 3} H_{\alpha \beta \ga}
H^{\alpha \beta \ga} -
e^{-2\phi} F_{\alpha \beta} F^{\alpha \beta} \right ] + I_{matter},
\label{ac}
\ee
where the strength of the gauge fields is described by
$F_{\mu \nu} = 2\na_{[\mu} A_{\nu]}$ and
the three index antisymmetric tensor is defined by the relation
\be
H_{\alpha \beta \ga} = \na_{\alpha}B_{\beta \ga} - A_{\alpha}F_{\beta
\ga} + cyclic .
\ee
In four dimensions $H_{\alpha \beta \ga}$ is equivalent to the
Peccei-Quin pseudo-scalar and implies the following:
\be
H_{\alpha \beta \ga} = {1 \over 2}\ep_{\alpha \beta \ga \delta}
e^{4 \phi} \na^{\delta} a.
\label{def}
\ee
A straightforward consequence of the 
definition (\ref{def}) is that the action (\ref{ac}) can be written
as
\be
I = \int d^4 x \sqrt{-g} 
\left [	- R + 2(\na \phi)^{2} 
+ {1 \over 2} e^{4 \phi} (\na a)^2
- e^{-2\phi} 
F_{\alpha \beta} F^{\alpha \beta} - a F_{\mu \nu} \ast F_{\mu \nu}
\right ] + I_{matter},
\label{act}
\ee
where  $\ast F_{\mu \nu} = 
{1 \over 2} \ep_{\mu \nu \delta \rho} F^{\delta \rho}$ .
\par
By means of the two-component spinor notation, we define a
Maxwell spinor by the relation
\be
F_{M M'~ N N'} = \phi_{MN}~ \ep_{M' N'} + \bphi _{M' N'}~ \ep_{M N},
\ee
where
\be
\phi_{A B} = {1 \over 2} F_{AB C'}{}{}^{C'},
\qquad \bphi_{A'B'} = {1 \over 2} F_{C}{}{}^{C}{}_{A'B'}.
\ee
Equations of motion derived from the variational principle are given by
\ben
\na_{M}{}{}^{N'} \phi^{MN} &=& \na^{N}{}{}_{M'} \bphi^{M'N'} ,\\
\na_{M}{}{}^{N'} \left (
z \phi^{MN} \right ) - \na^{N}{}{}_{M'} \left (
\bz \bphi^{M'N'} \right )
&=& \cJ^{N' N}(matter) ,\\
\na_{\mu} \na^{\mu} \phi - {1 \over 2 } e^{4 \phi} (\na a)^2 - e^{2 \phi}
\left ( \phi_{MN} \phi^{MN} + \bphi_{M'N'} \bphi^{M'N'} \right ) &=& 0, \\
\na_{\mu} \na^{\mu} a + 4 \na_{\mu} \phi \na^{\mu} a + 2 i e^{- 4 \phi}
\left ( \bphi_{M'N'} \bphi^{M'N'} - \phi_{M N} \phi^{M N} \right ) &=& 0, \\
6 \Lambda \ep_{AB} \ep_{A'B'} + 2 \Phi_{ABA'B'} = T_{AA' BB'},
\een
where we introduced a complex {\it axi-dilaton}
$z = a + i e^{-2 \phi}$, while $\Lambda = {R \over 24}$
and $\Phi_{ABA'B'}$ is the curvature spinor (sometimes called
Ricci spinor). The energy momentum tensor 
$T_{\mu \nu} = {2 \delta I \over \sqrt{-g} \delta g^{\mu \nu}}$
can be written as
\ben
T_{M M'~ N N'}(F, \phi, a) = 8 e^{-2 \phi} 
\phi_{M N}~ \bphi_{M' N'} &-& \ep_{M N}~ \ep_{M' N'}
\left [
2 (\na \phi )^2 + {1 \over 2} e^{4 \phi} ( \na a )^2 \right ] \\ \nonumber
&+&
4  \na_{M M'} \phi~ \na_{N N'} \phi 
+ e^{4 \phi}\na_{M M'} a~ \na_{N N'} a,
\een
while $\cJ^{\mu \nu}(matter) = {\delta I(matter) \over 4 \delta A_{\mu}}$.
\par
To begin with we define
the supercovariant derivatives for two-component spinors
$(\alpha^{A}{}{}_{(i)}, \beta_{A' (i)})$ \cite{rog1,rog2}
as follows:
\be
\hna_{M M'}~ \alpha_{K (i)} = \na_{M M'}~ \alpha_{K (i)} +
{i \over 2} e^{2 \phi} \na_{M M'} a~ \alpha_{K (i)}
+ \sqrt{2} e^{-\phi} \phi_{K M}~ a_{(ij)}~ \beta_{M'}^{(j)},
\label{d1}
\ee
\be
\hna_{M M'}~ \beta_{A' (i)} = \na_{M M'}~ \beta_{A' (i)} +
{i \over 2} e^{2 \phi} \na_{M M'} a ~\beta_{A' (i)}
- \sqrt{2} e^{-\phi} \bphi_{A' M'}~ a_{(ij)}~ \alpha_{M}^{(j)}.
\label{d2}
\ee
The {\it hatted derivative} (\ref{d1}-\ref{d2}) may be regarded as a
supersymmetry transformation about non-trivial gravitational, scalar
and $U(1)$ backgrounds.  The two-dimensional antisymmetric
matrices $a_{(ij)}$ where $i, j =1,2$ and $a_{(ij)} a^{(jm)} = - \delta_{i}^{m}$
are of the form 
$a_{(ij)} = \pmatrix{0&1\cr -1&0\cr}$. They constitute
two-dimensional equivalents of matrices introduced in \cite{cre78}.
\par
As was mentioned the EMAD gravity theory is a truncation
of $N = 4,~d = 4$ supergravity with only one vector field. Having this in mind
we shall keep another Latin index in the spinor transformation rules to make 
transparent linkage to the spinor transformations in the underlying
supergravity theory (see \cite{ka}-\cite{cre78} and references therein).
\par
For spinor fields
$(\alpha^{A}{}{}_{(i)}, \beta_{A' (i)})$ we 
describe the quantities which will be useful for our further considerations 
and to state  our results, namely
we define the Nester-Witten
two-form \cite{pen2} by
\be
\Theta (\alpha_{A}, \balpha_{A'}, \bbeta_{A}, \beta_{A'}) 
= - i \left (
\balpha_{B'(i)} \hna_{a}\alpha_{B}^{(i)}~dx^{a} \wedge dx^{b} +
\beta_{B'(i)} \hna_{a} \bbeta_{B }^{(i)}~dx^{a} \wedge dx^{b}
\right ).
\ee
Now, let $\Sigma$ be a spacelike two-surface spanned by
a three-dimensional compact hypersurface $\cal H$, with the unit
normal $t^{A A'}$ to $\cal H$. Then, using the Stokes'
theorem, after tedious calculations one obtains
\ben \label{dd}
\int_{\Sigma} \Theta (\alpha_{A}, \balpha_{A'}, \bbeta_{A}, \beta_{A'}) 
&=& \int_{\cal H}
d \Theta (\alpha_{A}, \balpha_{A'}, \bbeta_{A}, \beta_{A'}) \\ \nonumber
 &=& 
\int_{\cal H} d {\cal H} \bigg[
- \bigg (
\hD_{b} \alpha_{A(i)} \hD^{b} \balpha_{A'}^{(i)} + \hD_{b} \bbeta_{A(i)} \hD^{b} 
\beta_{A'}^{(i)}
\bigg) t^{A A'} +  {1 \over 2} T_{ab}(matter) \xi^{a} t^{b}
\\ \nonumber
&+&
\bigg(
(\delta \la_{A (i)}^{(\alpha)})^{\dagger} (\delta \la_{A'}^{(\alpha)(i)})
+ (\delta \la_{A (i)}^{(\beta)})^{\dagger} (\delta \la_{A'}^{(\beta)(i)})
\bigg) t^{A A'} \\ \nonumber
&+&
\bigg(
- \sqrt{2} e^{\phi} J_{AA'}({\bar z}, F) \alpha^{K(i)} \bbeta_{K(i)} + 
c.c. \bigg) t^{A A'} \\ \nonumber
&-& 2 \sqrt{2} J_{AA'}(\phi, F) \balpha^{K'(i)} \beta_{K'(i)} t^{A A'}
\bigg],
\een
where $\xi^{BB'} = \alpha^{B}_{(i)}~\balpha^{B'(i)} + 
\bbeta^{B~}_{(i)}~\beta^{B' (i)}$ is the future-directed null vector on 
an asymptotically flat spacetime \cite{bmas} while
$\hD_{a} = \hna_{a} - t_{a}t^{b} \hna_{b}$ is the 
projection of the supercovariant derivative into $\cal H$. We choose
the initial data for the Weyl equations $\hna_{AA'} \alpha^{A} = 0$ and
$\hna_{AA'} \bbeta^{A} = 0$ such that on $\cal H$ one has the following: 
\be
\hD_{A' A} \alpha^{A}{}{}_{(i)} = 0, \qquad
\hD_{A' A} \bbeta^{A}{}{}_{(i)} = 0.
\label{c3}
\ee 
The Witten's-like equations (\ref{c3}) constitute the first order elliptic
equations on $\alpha^{A}{}{}_{(i)}$ and $\bbeta^{A}{}{}_{(i)}$.
The energy momentum tensor $T_{ab}(matter)$ is defined by the relation
\be
T_{ab}(matter) = T_{ab}(total) - T_{ab}(F, \phi, a).
\ee
In relation (\ref{dd}) we have also defined the following complex currents:
\be
J_{A A'}(F, \phi) = \na_{A' B} (e^{- \phi} \phi^{B}{}{}_{A}),
\ee
and
\be
J_{AA'}({\bar z}, F) = \na_{A'B}( {\bar z} \phi_{A}{}{}^{B}).
\ee
Moreover, 
the adequate quantities appearing in equation (\ref{dd}) yield
\be
\delta \la_{A' (i)}^{(\alpha)} = 
\sqrt{2} \na_{A'B} \phi~ \alpha_{(i)}{}{}^{B} +
{i \over \sqrt{2}}e^{2 \phi}  \na_{A'}{}{}^{B} a~ \alpha_{{B}~(i)}
- 2 i e^{- \phi} \bphi_{ A'}{}{}^{C'}~ a_{(ij)}~ \beta_{C'}{}{}^{(j)},
\label{l1}
\ee
\be
\delta \la_{A (i)}^{(\beta)} =
\sqrt{2} \na_{A}{}{}^{B'} \phi~ \beta_{B' (i)} +
{i \over \sqrt{2}}e^{2 \phi}  \na_{A}{}{}^{B'} a~ \beta_{B' (i)}
+ 2 i e^{- \phi} \phi_{ A}{}{}^{C}~ a_{(ij)}~ \alpha_{C (j)}.
\label{l2}
\ee
The motivation for the specific factors that we used in the above definitions was to
derive the desired energy bound as well as to have the supergravity
transformation laws of the appropriate particles in the associated 
supergravity model. As was mentioned our model, EMAD gravity, constitutes the 
bosonic part of $N = 4,~d = 4$
supergravity with only one gauge field.
\par
In order to achieve the right-hand side of equation (\ref{dd})
positive one should have the following conditions satisfied:
\be
T_{ab}(matter) \xi^{a} t^{b} \ge
\bigg( \sqrt{2} e^{\phi} J_{AA'}({\bar z}, F) \alpha^{K (i)} \bbeta_{K (i)} + 
c.c. \bigg) t^{A A'} 
+ 2 \sqrt{2} J_{AA'}(\phi, F) \balpha^{K'(i)} \beta_{K'(i)} t^{A A'}.
\label{en1}
\ee
Condition (\ref{en1}) is stronger than the dominant energy condition
normally assumed in general relativity to prove the positivity of energy.
It reveals the fact that the local energy density is greater or equal to
the densities of the adequate charge densities. 

%%%%%%%%%%%%%%%%%%%%%%%%%%%%%%%%%%%%%
\par
On the spacelike two-surface $\Si$
we define 
a spinor basis
$(o_{A}, i^{A})$ on $\Si$ in such a manner
that $l^{a} = o^{A} \bo^{A'}$ and
$n^{a} = i^{A} \bi^{A}$ are future-directed outgoing and incoming null
normals to the
considered spacelike two-surface, while $m^{a} = o^{A} \bi^{A'}$ is tangent
vector to $\Si$. The differential operators $\delta = o^{A} \bi^{A'}
\hna_{A A'}$ and $\bar \delta = i^{A} \bo^{A'} \hna_{A A'}$ and
their generalization $\eth$ and $\eth'$ are all intrinsic on $\Si$
\cite{pen2}. \\
One
supposes further, that the number of spinor fields on $\Si$ is restricted so that
the space of such fields is isomorphic to a two-dimensional
complex vector space $\cal S$. Following Dougan and Mason \cite{mas} we can define
the two-dimensional 
complex vector space $\cal S$ by choosing the propagation 
equations for the spinor fields $\alpha_{A (i)}$ and $\bbeta_{A (i)}$
on the spacelike two-surface $\Si$. Namely, we shall consider spinor fields 
satisfying equations
$\eth \alpha _{A} = 0$,
(or $\eth' \balpha_{A'} = 0$) and $\eth \bbeta_{A} = 0$,
(or $\eth' \beta_{A'} = 0$).\\
If $\alpha_{A}^{\ab}$ and 
$\beta_{A'}^{\apb}$ where $\ab = 0, 1$ and  $\apb = 0', 1'$ are such two
fields, then
in the standard way \cite{pen2,ber1,mas} one may define a four-momentum 
corresponding to the spacelike two-surface $\Si$ by the following expression:
\be
P^{\ab \apb}(\Si, \alpha_{A}, \balpha_{A'}, \bbeta_{A}, \beta_{A'})
= \int_{\Si} \Theta
(\alpha_{A}^{\ab}, \balpha_{A'}^{\apb},\bbeta_{A}^{\ab}, \beta_{A'}^{\apb}),
\label{qua}
\ee
where $\Si$ is normally assumed to be 
homeomorphic to the two-sphere.
\par
When the hypersurface $\Si$ is a cross section of null infinity and when
$(\alpha_{A}^{\ab}{}{}_{(\infty)}, \beta_{A'}^{\apb}{}{}_{(\infty)})$
are required to be asymptotically constant spinors, then 
$P^{\ab \apb}
(\Si, \alpha_{A}, \balpha_{A'}, \bbeta_{A}, \beta_{A'})$
is the Bondi four-momentum. On the other hand, at spacelike infinity
one gets the ADM four-momentum.
\par
From now on (for simplicity) we shall 
drop the underlined indices and
denote $P^{0 0'} = P$
and $\alpha_{A(i)}^{0} = \alpha_{A(i)}$, etc..
The Geroch-Held-Penrose formalism \cite{pen,ger} 
enables us to express
the quantity
$P(\Si, \alpha_{A}, \balpha_{A'}, \bbeta_{A}, \beta_{A'})$ 
in the following form:
\ben \label{pp}
P(\Si, \alpha_{A}, \balpha_{A'}, \bbeta_{A}, \beta_{A'})
= \\ \nonumber
 \int_{\Si} d\Si~ \bigg[ 
- \balpha_{0'}^{(i)} \bigg (
\eth \alpha_{1 (i)} + \rho' \alpha_{0 (i)} + {i e^{2 \phi} \over 2}
\eth a~ \alpha_{1 (i)} + \sqrt{2} e^{- \phi}
\phi_{(1)}~a_{(ij)}~\beta_{1'}^{(j)} \bigg ) \\ \nonumber
+ \balpha_{1'}^{(i)} \bigg (
\eth' \alpha_{0 (i)} + \rho~ \alpha_{1 (i)} + {i e^{2 \phi} \over 2}
\eth' a~ \alpha_{0 (i)} + \sqrt{2} e^{- \phi}
\phi_{(1)}~a_{(ij)}~\beta_{0'}^{(j)} \bigg ) \bigg] \\ \nonumber
+ \int_{\Si} d\Si~
\bigg[
- \beta_{0'}^{(i)} \bigg (
\eth \bbeta_{1 (i)} + \rho' \bbeta_{0 (i)} - {i e^{2 \phi} \over 2}
\eth a~ \bbeta_{1 (i)} - \sqrt{2} e^{- \phi}
\phi_{(1)}~a_{(ij)}~\balpha_{1'}^{(j)} \bigg ) \\ \nonumber
+ \beta_{1'}^{(i)} \bigg (
\eth' \bbeta_{0 (i)} + \rho~ \bbeta_{1 (i)} - {i e^{2 \phi} \over 2}
\eth' a~ \bbeta_{0 (i)} - \sqrt{2} e^{- \phi}
\phi_{(1)}~a_{(ij)}~\balpha_{0'}^{(j)} \bigg ) \bigg],
\een
where 
$\phi_{(1)} = \phi_{AB} o^{A} i^{B}$ is a complex scalar
generated from a symmetric two-spinor describing the Maxwell field,
and
$\alpha_{0 (i)} = \alpha_{A (i)} o^{A},~~\alpha_{1 (i)} =
\alpha_{A (i)} i^{A}$
while $\bbeta_{0 (i)} = \bbeta_{A (i)} o^{A}, 
~~\bbeta_{1 (i)} = \bbeta_{A (i)} i^{A}$. 
\par
Thus we shall
consider the following propagation equations on $\Si$:
\ben \label{u1}
\bi^{A'} \bm^{b} \hna_{b} \beta_{A'(i)} =
\eth' \beta_{1'(i)} + \brho'~ \beta_{0'(i)} &+& {i e^{2 \phi} \over 2}
\eth' a~\beta_{1'(i)} \\ \nonumber
&-& \sqrt{2} e^{- \phi} \bphi_{(1')}~a_{(ij)} 
\alpha_{1}^{(j)} = 0,
\een
\be
i^{A} m^{b} \hna_{b} \alpha_{A(i)} =
\eth \alpha_{1 (i)} + \rho'~ \alpha_{0 (i)} + {i e^{2 \phi} \over 2}
\eth a~ \alpha_{1 (i)} + \sqrt{2} e^{- \phi} \phi_{(1)} a_{(ij)}~
\beta_{1'}^{(j)} = 0,
\label{w2}
\ee
and their conjugates.
It will be easily verified that by virtue of
the equations (\ref{u1}) and (\ref{w2})  
and by means of integration by parts we reach to the expression
\ben \label{kkk}
P(\Si, \alpha_{A}, \balpha_{A'}, \bbeta_{A}, \beta_{A'}) &=& 
\int_{\Si} d\Si \bigg( 
\rho' \balpha_{0'(i)}~\alpha_{0}^{(i)} + \rho~ \balpha_{1'(i)}
~\alpha_{1}^{(i)} \\ \nonumber
&+& \sqrt{2} e^{- \phi} \bphi_{(1')} a_{(ij)}~\alpha_{0}^{(i)}~
\bbeta_{1}^{(j)} +
\sqrt{2} e^{- \phi} \phi_{(1)}~\balpha_{1'}^{(i)}~a_{(ij)}
~\beta_{0'}^{(j)}
\bigg)  \\ \nonumber
&+& \int_{\Si} d\Si \bigg( 
\rho' \beta_{0'(i)}~\bbeta_{0}^{(i)} + \rho~ \beta_{1'(i)}
~\bbeta_{1}^{(i)} \\ \nonumber
&-&
\sqrt{2} e^{- \phi} \bphi_{(1')} a_{(ij)}~\alpha_{1}^{(i)}~
\bbeta_{0}^{(j)} -
\sqrt{2} e^{- \phi} \phi_{(1)}~\balpha_{0'}^{(i)}~a_{(ij)}
~\beta_{1'}^{(j)}
\bigg).
\een
For two-spinors $\mu_{A}, \bmu_{A'}, \bbe_{A}, \eta_{A'}$ on ${\cal H}$
satisfying the Witten equations (\ref{c3}) and the condition
(\ref{en1}) we have
$P(\Si,\mu_{A}, \bmu_{A'}, \bbe_{A'}, \eta_{A}) \ge 0$.
Having given  
$\alpha_{A (i)}$ and $\beta_{A' (i)}$ and their conjugates satisfying
relations
(\ref{u1}-\ref{w2}) one can find such $\mu_{1(i)}$ and 
$\bbe_{1(i)}$
on ${\cal H}$ fulfilling the boundary conditions 
$\mu_{1(i)} = \alpha_{1(i)}$
and $\bbe_{1(i)}  = \bbeta_{1(i)}$ on $\Si$.
The existence of such $\mu_{1(i)}$ and $\bbe_{1(i)}$ can be proved 
\cite{mas}
by
the direct application of the Fredholm alternative (see,
e.g., \cite{gil}) or by using the known theorems \cite{hor} 
for compact hypersurfaces.
Having in mind (\ref{kkk}), we shall relate 
$P(\Si, \alpha_{A}, \balpha_{A'}, \bbeta_{A}, \beta_{A'})$
to $P(\Si,\mu_{A}, \bmu_{A'}, \bbe_{A'}, \eta_{A})$.
First one replaces $\alpha_{1(i)}$ by $\mu_{1(i)}$ 
and $\bbeta_{1(i)}$ by 
$\bbe_{1(i)}$ in equation (\ref{kkk}). Then, we use
equations (\ref{u1}) and (\ref{w2}) and integration by parts.
Finally one gets the formula
\ben
P(\Si,\mu_{A}, \bmu_{A'}, \bbe_{A}, \eta_{A'}) &=&   
P(\Si, \alpha_{A}, \balpha_{A'}, \bbeta_{A}, \beta_{A'}) \\ \nonumber
 &-&
\int_{\Si} d \Si~ \rho' \mid \alpha_{0 (i)} - \mu_{0 (i)} \mid^2
- \int_{\Si} d \Si~ \rho' \mid \bbeta_{0 (i)} - \bbe_{0 (i)} \mid^2.
\label{ddd}
\een
We assume that $\rho' \ge 0$ on the spacelike two-surface $\Si$
and because of this fact
the left-hand side of the above equation is greater or equal to zero 
on $\Si$, namely
\be
P(\Si, \alpha_{A}, \balpha_{A'}, \bbeta_{A}, \beta_{A'}) \ge 0.
\ee
Let us remark that the case of
the other possible equations of spinors propagation of the forms
\ben \label{b1}
\bo^{A'} m^{b} \hna_{b} \beta_{A'(i)} =
\eth \beta_{0'(i)} + \brho~ \beta_{1'(i)} &+& {i e^{2 \phi} \over 2}
\eth a~\beta_{0'(i)} \\ \nonumber
&-& \sqrt{2} e^{- \phi} \bphi_{(1')}~a_{(ij)} 
\alpha_{0}^{(j)} = 0,
\een
\be
o^{A} \bm^{b} \hna_{b}\alpha_{A(i)} =
\eth' \alpha_{0 (i)} + \rho~ \alpha_{1 (i)} + {i e^{2 \phi} \over 2}
\eth' a~ \alpha_{0 (i)} + \sqrt{2} e^{- \phi} \phi_{(1)} a_{(ij)}~
\beta_{0'}^{(j)} = 0,
\label{b2}
\ee 
and their conjugates, 
follow almost in the same way 
with a slight modification of the arguments due to the fact that $\rho = \brho$
and the assumption
$\rho \le 0$.
All these lead us to the same conclusion.
Thus, from the above considerations we can deduce the following result:\\
\noindent
{\bf Theorem}\\
Consider a three-dimensional compact spacelike hypersurface ${\cal H}$
on which the condition (\ref{en1})
is satisfied.
Let $\Si$ be 
a spacelike two-surface
spanned by a three-dimensional hypersurface
$\cal H$~$(\Si = \p {\cal H})$. Suppose further, that 
the propagation equations
(\ref{u1}-\ref{w2}) hold on $\Si$
and assume that $\rho' \ge 0$ (or $\rho \le 0$ in the case of 
equations (\ref{b1})-(\ref{b2})).
Then, it follows\\
$\int_{\Si}~\Theta (\alpha_{A}, \balpha_{A'}, 
\bbeta_{A}, \beta_{A'}) \ge 0$. \\

\vspace{0.02cm}
Having established the fact that 
$P(\Si, \alpha_{A}, \balpha_{A'}, \bbeta_{A}, \beta_{A'}) \ge 0$
we can proceed to the proof of the positivity of energy in EMAD
gravity. 
Thus, let us 
consider a three-surface extension inwards from infinity (spacelike
or null) and foliate it by two-surfaces $\Si_{r}$ ~$(r
\rightarrow \infty$ at infinity). \\
We notice that one can always find families of two-surfaces approaching
spacelike or null infinity with $\rho' > 0$. 
Namely, following Chrusciel \cite{chr} we shall call a hypersurface 
asymptotically flat if it contains an asymptotically flat end, i.e., a data set
$(\Sigma_{end}, g_{ij}, K_{ij})$ with gauge and dilaton fields such that 
$\Sigma_{end}$ is diffeomorphic to ${\bf R}^3$ minus a ball and the 
following asymptotic conditions are fulfilled:
\ben
\vert g_{ij}  - \delta_{ij} \vert + r \vert \p_{a}g_{ij} \vert
+ ... + r^k \vert \p_{a_{1}...a_{k}}g_{ij} \vert +
r \vert K_{ij} \vert + ... + r^k \vert \p_{a_{1}...a_{k}}K_{ij} \vert
\le {\cal O}\bigg( {1\over r} \bigg), \\
\vert F_{\alpha \beta} \vert + r \vert \p_{a} F_{\alpha \beta} \vert
+ ... + r^k \vert \p_{a_{1}...a_{k}}F_{\alpha \beta} \vert
\le {\cal O}\bigg( {1 \over r^2} \bigg), \\
\phi = \phi_{\infty} + {\cal O}\bigg( {1\over r} \bigg).
\een
\par
In a flat spacetime $\rho' = {1 \over r}$
for spheres around $r = 0$.
Having in mind the results presented in \cite{asymp}
one can deduce that in asymptotically flat case near
infinity we have $\rho'= {1 \over r} +  {\cal O} ({1 \over r^2})$. 
The same situation 
takes place for cross sections of a null hypersurface approaching
a given cross section of future null infinity \cite{pen2}.\\
On $\Si_{r}$ we assume that
$\rho' \ge 0$ for $r$ large enough. Suppose further that we have given
values of asymptotically constant spinors 
$(\alpha_{A (\infty)}, \beta_{A'(\infty)})$ at infinity. Let us extend
the components of $\alpha_{1 (i)}$ and $\bbeta_{1 (i)}$
to all hypersurfaces $\Si_{r}$ in an arbitrary differentiable way. 
Next define
$\alpha_{0(i)}$ and $\bbeta_{0 (i)}$ such that
\be
\alpha_{0(i)} = {1 \over \rho'}
\left (
- \eth \alpha_{1 (i)} - {i e^{2 \phi} \over 2}
\eth a~ \alpha_{1 (i)} - \sqrt{2} e^{- \phi} \phi_{(1)} a_{(ij)}~
\beta_{1'}^{(j)} \right ),
\ee
and
\be
\bbeta_{0(i)} = {1 \over \rho'}
\left (
- \eth \bbeta_{1 (i)} + {i e^{2 \phi} \over 2}
\eth a~ \bbeta_{1 (i)} + \sqrt{2} e^{- \phi} \phi_{(1)} a_{(ij)}~
\balpha_{1'}^{(j)} \right ).
\ee
Hence, from the previous
theorem we have that
$\int_{\Si_{r}}~ 
\Theta (\alpha_{A}, \balpha_{A'}, 
\bbeta_{A}, \beta_{A'}) \ge 0$ for all $r > R$. 
In an asymptotically flat spacetime we have introduced ${\Si_{r}} $
a family of two-surfaces approaching spacelike infinity or a cross section
of future null infinity as $r \rightarrow \infty$. For asymptotically
constant spinors $(\alpha_{A (\infty)}, \beta_{A'(\infty)})$ at spacelike
or null infinity we can define
\be
P_{i} k^{i} = \lim\limits_{r \rightarrow \infty} 
\int_{\Si_{r}}~ 
\Theta (\alpha_{A}, \balpha_{A'}, \bbeta_{A}, \beta_{A'})
\ge 0,
\ee
which is respectively the ADM or Bondi momentum, while 
$k^{a} = \lim\limits_{r \rightarrow \infty}
\left ( \alpha^{A}_{(i)}~\balpha^{A'(i)} + 
\bbeta^{A~}_{(i)}~\beta^{A' (i)} \right )$
is a future pointing asymptotic null translation.
Using a
linear combinations of null vectors, the quantity
$P_{i}$ can be determined completely.\\
Then, in order to prove the positivity of energy  it is necessary to get
the relation that $P_{i} u^{i} \ge 0$, for all
future pointed asymptotic translations $u^{i}$. On the other hand,
since a translation $u^{i}$ can be expressed as a linear combination
of two future pointed null
vectors, it will be sufficient to find that
$P_{i}k^{i} \ge 0$, for all null $k^{i}$.\\
One can notice that the above limit exist because it is energy.
In the above proof spinor fields defining the energy need to be 
defined near infinity and obey the propagation 
equations (\ref{u1}
-\ref{w2}).
Positivity of it is proved by having in
mind the previous theorem and
the Witten's propagation equation 
for auxiliary spinor fields $\mu_{A(i)}$ and $\bbe_{A(i)}$ 
on a compact set.\\
We can therefore assert the following estimate:\\
{\bf Theorem}\\
Suppose that the condition (\ref{en1})
is satisfied in an asymptotically flat
spacetime. Then, $P_{i}k^{i} \ge 0$, where $P_{i}$ is the ADM 
or Bondi momentum.

%%%%%%%%%%%%%%%%%%%%%%%%%%%%%%%%%%%%%%%%%%%%%%%%%%%%%%%%%%%%%%%
\vspace{0.2cm}
Now we take up a question of the stronger inequality the ADM or Bondi
mass should satisfy. In order to do so one considers the situation when
$ \alpha^{A(i)}$ and $\bbeta_{A(i)}$ approach the constant 
spinors at infinity.
Using the exact form of the two-surface bivector
$dS_{ab} = l_{[a}n_{b]}dS$, one obtains
\ben \label{ineq}
\int_{\Si} \Theta
(\alpha_{A}^{\ab}, \balpha_{A'}^{\apb},\bbeta_{A}^{\ab}, \beta_{A'}^{\apb})
= \tP^{m}(\alpha_{A}^{\ab}, \balpha_{A'}^{\apb}, \bbeta_{A}^{\ab}, \beta_{A'}^{\apb})k_{m} 
&+&
{1 \over \sqrt{2}} Q_{(F-\phi)}
\bigg( \balpha^{A'(i)}_{(\infty)}\beta_{A'(i)(\infty)} \bigg) \\ \nonumber
&+&
{1 \over \sqrt{2}} P_{(F-\phi)}
\bigg( \balpha^{A'(i)}_{(\infty)}\beta_{A'(i)(\infty)} \bigg),
\een
where the {\it dilaton-electric} charge is defined as \cite{rog2}
\be
Q_{(F-\phi)} = 2\int_{S^{\infty}} dS e^{-\phi_{\infty}}~ Re~ \phi_{(1)},
\ee
and consequently the {\it magnetic-dilaton} charge is expressed as follows:
\be
P_{(F-\phi)} = 2\int_{S^{\infty}} dS e^{-\phi_{\infty}}~ Im~ \phi_{(1)},
\ee
while $\tP_{i} k^{i}$ is denoted by the expression
\be
\tP^{i} k_{i} = \lim\limits_{r \rightarrow \infty} 
\int_{\Si_{r}}~ 
\tTheta (\alpha_{A}^{\ab}, \balpha_{A'}^{\apb}, \bbeta_{A}^{\ab}, \beta_{A'}^{\apb}).
\label{ene}
\ee
In equation (\ref{ene}) we have used the following definition:
\be
\tTheta (\alpha_{A}, \balpha_{A'}, \bbeta_{A}, \beta_{A'})
= - i \left (
\balpha_{B'(i)} \na_{a}\alpha_{B}^{(i)}~dx^{a} \wedge dx^{b} +
\beta_{B'(i)} \na_{a} \bbeta_{B }^{(i)}~dx^{a} \wedge dx^{b}
\right ).
\ee
Next following \cite{mas} let us define quantity
$m(\Si)$ of the form
\be
m^{2}(\Si) = 
\tP^{\ab \apb}(\Si, \alpha_{A}, \balpha_{A'}, \bbeta_{A}, \beta_{A'})~
\tP^{\bb \bbb}(\Si, \alpha_{A}, \balpha_{A'}, \bbeta_{A}, \beta_{A'})
\ep_{\ab \bb}~ \ep_{\apb \bbb},
\ee
where $\ep_{\ab \bb} = \bigg( \alpha^{A}_{\ab}~\alpha^{B}_{\bb} + 
\bbeta^{A}_{\ab}~\bbeta^{B}_{\bb} \bigg) \ep_{AB}.$ From the propagation
equations for spinors, i.e., 
$\delta \alpha_{\ab}^{A} = \delta \bbeta_{\ab}^{A} = 0$ or
$\bdelta \balpha_{\apb}^{A'} = \bdelta \beta_{\apb}^{A'} = 0$
one has that $\delta \ep_{\ab \bb} = 0$ (or $\bdelta \ep_{\apb \bbb} = 0$).
So that $\ep_{\ab \bb}$ is anti-holomorphic ($\ep_{\apb \bbb}$ is holomorphic)
and global function on the sphere and by means of the Liouville's
theorem it is constant.
We shall call $m(\Sigma)$ the mass.
If $\Sigma$ is a cross section of null infinity we shall have the Bondi mass,
while at spatial infinity one gets the ADM mass. Moreover one should demand that the 
two-dimensional complex vector space will be the space of asymptotically constant spinors.
\par 
Thus, from equation (\ref{ineq}) we deduce that the mass $m(\Sigma)$ must be
positive for all spinors $\balpha^{A'(i)}_{(\infty)}$ and $\beta_{A'(i)(\infty)}$,
and the following inequality binding mass with {\it dilaton-electric} and
{\it dilaton-magnetic} charges should be fulfilled:
\be
m^2(\Sigma) \ge Q_{(F-\phi)}^2 + P_{(F-\phi)} ^2.
\ee

%%%%%%%%%%%%%%%%%%%%%%%%%%%%%%%%%%%%%%%%%%%%%%%%%%%%%%%%%%%%%%%%%%
\section{Conclusions}
In our work we considered the positivity of the ADM and Bondi energy in
EMAD gravity. Extending the reasoning presented in \cite{ber,mas}
we gave the proof of positivity of energy without proving the
existence of asymptotically constant spinors satisfying differential
equations on a non-compact hypersurface.
The spinor fields defining the energy $(\alpha_{A(i)}, \beta_{A'(i)})$
need only to be defined near infinity (near spacelike infinity to get
the ADM energy or near null infinity to achieve the Bondi energy).
They also should fulfill propagation 
equations on a spacelike two-surface $\Si$ and the
condition (\ref{en1}) for $T_{ab}(matter)$ and fields in
the theory under consideration. 
We have also establish the stronger inequality binding the ADM and Bondi mass
with the {\it dilaton-electric} and {\it dilaton-magnetic} charges.
However, our new proof of the positive
energy theorem does not simplify the proof for the flat spacetime. We hope
to return to this problem elsewhere.

\vspace{0.5cm}
%%%%%%%%%%%%%%%%%%%%%%%%%%%%%%%%%%%%%%%%%%%%%%%%%%%%%%%%%%%%%%%%%%%%%%%%
\noindent
{\bf Acknowledgements:}\\
I would like to thank the unknown referees for very useful comments.
%MR was supported in part by KBN grant.

%%%%%%%%%%%%%%%%%%%%%%%%%%%%%%%%%%%%%%%%%%%%%%%%%%%%%%%%%%%%%%%%%%%

%\end{references}
\end{document}